# Electrically Driven Hyperbolic Nanophotonic Resonators as High Speed, Spectrally Selective Thermal Radiators

*John Andris Roberts[1], Po-Hsun Ho[2], Shang-Jie Yu[2], and Jonathan A. Fan[2]\**

1. Department of Applied Physics, Stanford University, Stanford, CA 94305, U.S.A.//
2. Department of Electrical Engineering, Stanford University, Stanford, CA 94305, U.S.A.

Correspondence and requests for materials should be addressed to J.A.F. (jonfan@stanford.edu)

Abstract

We introduce and experimentally demonstrate a new class of electrically driven thermal emitter based on globally aligned carbon nanotube metamaterials patterned as nanoscale ribbons. The metamaterial ribbons exhibit electronic and photonic properties with extreme anisotropy, which enable low loss, wavelength-compressed hyperbolic photonic modes along one axis and high electrical resistivity and efficient Joule heating along the other axis. Devices batch-fabricated on a single chip emit linearly polarized thermal radiation with peak wavelengths dictated by their hyperbolic resonances, and their low thermal mass yields infrared radiation modulation rates as high as one megahertz. As a proof-of-concept demonstration, we show that two sets of thermal emitters on a single chip, each operating with different spectral peak positions and modulation rates, can be used to sense carbon dioxide with a single detector. We anticipate that the combination of batch fabrication, wide modulation bandwidth, and customized spectral tuning with hyperbolic chip-based thermal emitters will enable new modalities in multiplexed infrared sources for sensing, imaging, and metrology applications.

**Main Text**

Mid-infrared light sources are the basis for numerous applications in the areas of environmental monitoring, industrial safety, and security.[1] An ideal light source would combine low cost, a narrowband emission spectrum that coincides with a desired molecular resonance frequency, and a high modulation rate that enables frequency-multiplexed detection schemes. Advanced optoelectronic sources based on quantum cascade lasers and mid-infrared light emitting diodes offer fast modulation speeds, but their high cost precludes their use in many mainstream applications.[2] As such, the most common infrared sources in conventional systems are incandescent emitters that utilize an electrically heated filament for thermal emission.[3,4] Traditional incandescent sources are cheap to fabricate and package, but they emit light across a broad wavelength range and have modulation rates limited to approximately 10 Hz,[5] making them less than ideal in most sensing applications.

These fundamental limitations have motivated the development of thermal emitters with new capabilities.[6–8] In most cases to date, approaches have been proposed that enhance either spectral selectivity or modulation rate. Spectrally tailored thermal emitters have been made possible by globally heating metamaterials,[9–19] which are subwavelength-scale structured media that can support narrowly peaked emissivity spectra corresponding to engineered absorptive resonances, as dictated by Kirchhoff's law. Fast modulation from thermal emitters has been achieved by use of novel materials that feature novel tuning mechanisms[20–24] or exceptionally small volumes,[25–30] which enable fast variations in temperature.[31–33]

While it may appear straightforward to combine the two concepts above to achieve high speed, self-contained, spectrally selective thermal emitters with minimal heating volumes, there remain fundamental material roadblocks that make this a non-trivial proposition. For example, the refractory plasmonic materials typically used in thermal metamaterials have relatively low electrical resistivities, making it prohibitively difficult to design structures that are simultaneously efficient resistive heaters and optically resonant mid-infrared elements. Devices featuring fast thermal emission modulation also often utilize low dimensional materials that confine infrared photons to extremely small length scales, making it challenging to efficiently outcouple these photons to free space. Hybrid device schemes that couple thermal emitters to optical cavities can leverage the spectral selectivity and optical density of states of cavities to circumvent these material limitations.[25,27,32–34] However, they impose restrictions on the implementation of multiple emitters, each operating at unique wavelengths, and on the flexibility of integration on a chip.

We propose and demonstrate hyperbolic carbon nanotube (CNT) metamaterial nanoribbons as mid-infrared thermal emitters that function as both high-speed electrically driven heaters and optically resonant, spectrally selective nanostructures. A schematic of our device concept is shown in Figure 1a. A CNT metamaterial is prepared in bulk using vacuum filtration and patterned into nanoribbons that function as infrared hyperbolic metamaterial resonators capable of confining light to extreme, subwavelength-scale dimensions.[35,36] At the hyperbolic resonance frequency, the nanoribbon array has increased absorption, which translates to increased narrowband emissivity and subsequent spectrally-selective thermal emission upon heating. The nanoribbons are resistively heated by direct electrical biasing so that the CNT nanoribbons function as both optical hyperbolic resonators and heating elements.

Our use of hyperbolic CNT metamaterial nanostructures as incandescent emitters offers several unique advantages. First, CNT metamaterials are well-suited for heterogeneous integration onto chip-based platforms, as they can be readily transferred onto arbitrary substrates.[37–39] Second, the broadband hyperbolic dispersion of the metamaterial enables infrared-resonant CNT structures with nanoscale dimensions, leading to electrically-driven, low thermal mass, high speed emitters with tailorable emission wavelengths using a simple ribbon geometry.[35,36,40–45] While our hyperbolic resonators feature mode confinement in small volumes, it is not so extreme as to prevent the resonators from efficiently coupling to free space. Third, CNTs have excellent high temperature stability,[31,34,46–51] and the hyperbolic behavior of CNT metamaterials is known to persist at high temperatures and has been used to control the emissivity of a globally heated surface.[42] Fourth, the electrical resistivity perpendicular to the CNTs is around $2\text{-}3 \times 10^{-2}$ ohm·cm (see the Supporting Information), which is over two orders of magnitude higher than typical resistivities of refractory plasmonic metals[52] and allows relatively low currents to be used to achieve efficient heating. The much higher conductivity parallel to the CNTs, on the other hand, allows the nanoribbons to support high quality hyperbolic infrared resonances. In this way, the anisotropic macroscopic ordering of the self-assembled CNTs resolves the conflict between the electrical and optical requirements for a self-contained resonant heating element. Combining these advantages, our device concept enables multiple electrically-driven emitters, designed for different wavelengths and with fast modulation rates, to be fabricated on the same chip.

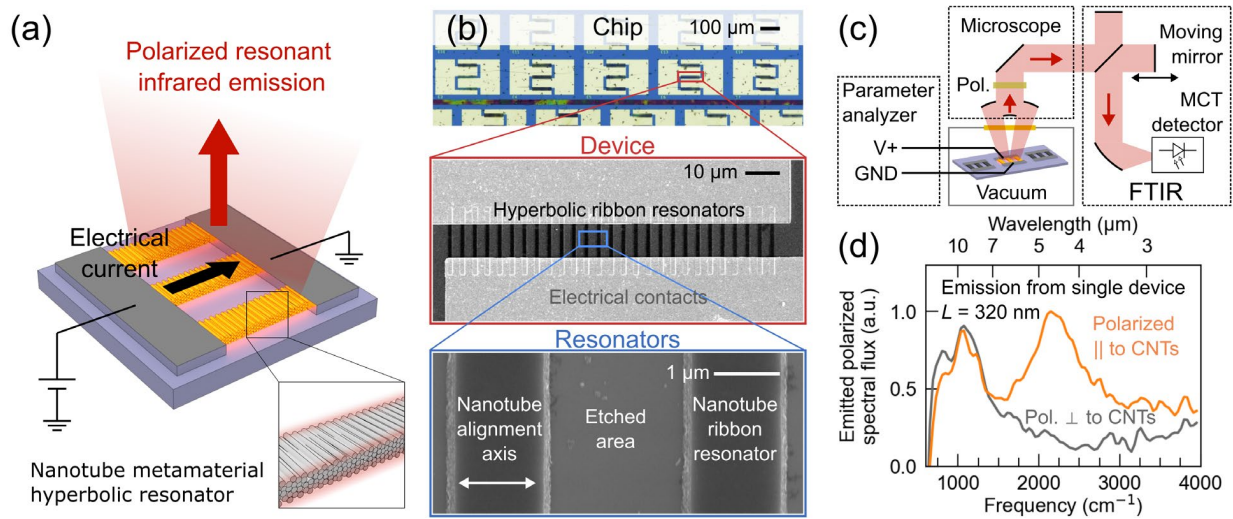

**Figure 1.** Device configuration and emission spectra of CNT metamaterial nanoribbons. (a) Device concept showing carbon nanotube (CNT) metamaterial ribbons functioning as both resistive heaters and hyperbolic mid-infrared resonators. (b) Top: Optical microscope image of a chip with several devices fabricated from a single globally-aligned CNT film on a thermally oxidized silicon substrate. Middle: Scanning electron micrograph (SEM) of a row of resonators in one device. Bottom: High magnification SEM of two resonators (darker areas), with the alignment direction of the CNTs indicated. (c) Schematic of the measurement apparatus. The sample is mounted in a high-vacuum chamber with a window, and a single device is electrically probed. Thermal emission from the single device is collected by a microscope objective and its spectrum is measured using the FTIR spectrometer. A polarizer can be rotated to analyze the polarization of emitted light. (d) Emission spectra from a single device with $L$ = 320 nm driven with 14 mA of current, polarized along (orange) and perpendicular to (gray) the CNT alignment axis. A polarized emission peak at wavelengths near 4.5 μm corresponds to thermal emission mediated by the hyperbolic plasmon mode. The spectra are normalized using the measured relative instrument response function, so that the result is a spectral radiance in arbitrary units (see the Supporting Information).

The fabricated devices on a silicon substrate are shown in Figure 1b. As shown in the top optical microscope image, many emitters can be patterned on one chip because the vacuum filtration process creates chip-scale CNT films with global alignment and consistent thickness. Each emitter can be designed to emit at a different resonant wavelength by varying the ribbon width, $L$, and angle relative to the CNT alignment axis, $\theta$, which alters the Fabry-Perot resonance condition of the hyperbolic waveguide modes in the ribbon.[35] Each device comprises three arrays of CNT ribbons with the same values of $L$ and $\theta$, which are connected in parallel using interdigitated metal electrodes. The interdigitated structure allows for relatively short ribbon lengths (10 μm), which reduces the likelihood of a given ribbon having a defect that severs its electrical conductivity, while maintaining a large enough total emitting area. Values of $L$ range between 320-900 nm, and the CNT film has a thickness of ~115 nm (see the Supporting Information).

We measure the infrared spectrum of emitted light from individual devices using a Fourier transform infrared (FTIR) spectrometer coupled to a microscope, depicted in Figure 1c. The device is kept under high vacuum while a bias is applied using a semiconductor parameter analyzer. For each device, we at first apply relatively high voltages, over an extended time, to 'burn out' initial visibly emitting hot spots and obtain more uniform visible emission from a larger number of remaining hot spots (see the Supporting Information). The light emitted by the device is collected by the microscope objective, passed through a polarizer, modulated by the scanning mirror of the FTIR, and received by the FTIR's internal detector. The polarizer is used because the CNT hyperbolic plasmon resonance is known to be excited by light polarized along the CNT alignment axis,[40,41,35] and we expect the resonant thermal emission to be polarized along this axis.

We perform automated, cycled emission measurements to obtain emission spectra for both polarizations (see the Supporting Information).

The measured emission spectra from these devices show clear emission peaks originating from the hyperbolic resonances of the metamaterial nanoribbons. Figure 1d shows the emission spectra for each polarization from a representative device with $L = 320$ nm and $\theta = 90°$ under a constant 14 mA bias, normalized using the measured spectral response function[53–55] of the spectrometer (see the Supporting Information). Two main spectral features are apparent. Near 4.5 µm, an emission peak appears that is polarized along the CNT alignment axis. This polarization signature is expected for the hyperbolic plasmon resonance, and the peak appears at a frequency near where the CNT metamaterials were previously observed to have hyperbolic dispersion.[35,36] An unpolarized emission peak at wavelengths near 10 µm can be attributed to a phonon resonance in the $SiO_2$ substrate[56] and indicates heating in the substrate.

To further probe the origin of the thermal emission resonance and its tunability, we repeat these emission measurements using devices with different ribbon widths and angles, all patterned from the same CNT film on one chip. The scaling trends we observe indicate that the polarized emission peak arises from the hyperbolic plasmon resonance in the nanotube metamaterial ribbons and demonstrate that multiple emitters with different peak wavelengths can be patterned in one step on the same device. Figure 2a shows the emission spectra of three devices, with the device from Figure 1c shown as Device B. We plot the difference in emitted spectral radiance between polarizations parallel and perpendicular to the CNT axis, $I_\parallel - I_\perp$, because the resonance of interest is polarized along the CNT alignment axis. Device A has wider ribbon resonators ($L = 680$ nm) than Device B, with a ribbon direction that is still perpendicular to the CNT alignment axis ($\theta = 90°$).

In the hyperbolic resonator model, we expect the resonance to be at a lower frequency for wider ribbons, as previously observed in far-field transmission measurements.[35,40] As expected, the resonance appears at a longer wavelength, with a peak near 5.5 µm. The nanoribbons in Device C are slightly wider than those in Device A, with $L = 410$ nm, but feature a CNT alignment angle of $\theta = 30°$. Peak thermal emission in these devices shifts to even longer wavelengths, with a maximum near 6.5 µm, because the resonant hyperbolic waveguide mode in the nanoribbon lies along a portion of the hyperbolic isofrequency surface featuring a large in-plane wavevector.[35]

A theoretical model of the emission spectra agrees well with our experimental findings and further reinforces that the hyperbolic metamaterial resonances control the emission spectra. We use the local Kirchhoff law for an anisothermal emitter because the CNT resonators are locally heated,[33,57] and we consider a simplistic model that assumes that the emission originates from CNT resonators at one temperature, $T_{CNT}$, and from the heated oxide at an arbitrarily assumed temperature $T_{ox}$. With these assumptions, the polarized emission spectral distribution in a direction normal to the surface of the chip can be modelled as:

$$I_{\parallel,\perp}(\omega) = I_{BB}(T_{CNT}, \omega) A_{CNT}(\omega) + I_{BB}(T_{oxide}, \omega) A_{oxide}(\omega)$$

$I_{BB}(T, \omega)$ is the blackbody spectral radiance and $A_{CNT}(\omega)$ and $A_{oxide}(\omega)$ are the integrated absorption of a normally-incident plane wave in the CNT resonator and oxide, respectively. The first term gives the resonator absorption multiplied by a Planck spectrum at the CNT temperature (estimated using visible emission spectroscopy as discussed below), while the second term gives the same for the heated oxide in the substrate. We model the integrated absorption terms using a two-dimensional finite element method electromagnetic simulation (see the Supporting Information) with previously-measured effective dielectric functions for the CNT metamaterial.[35,36] Our model gives emission spectra that agree remarkably well with the experimentally measured spectra (Figure 2b). The polarized emission peak originates from a resonance in the CNTs with a spectral shape and simulated field profiles (Figure 2b, right) consistent with earlier modeling of the CNT hyperbolic resonance.[35,43] The agreement between this model and the experimentally observed emission spectra highlights the stability of the hyperbolic resonance at high temperatures.

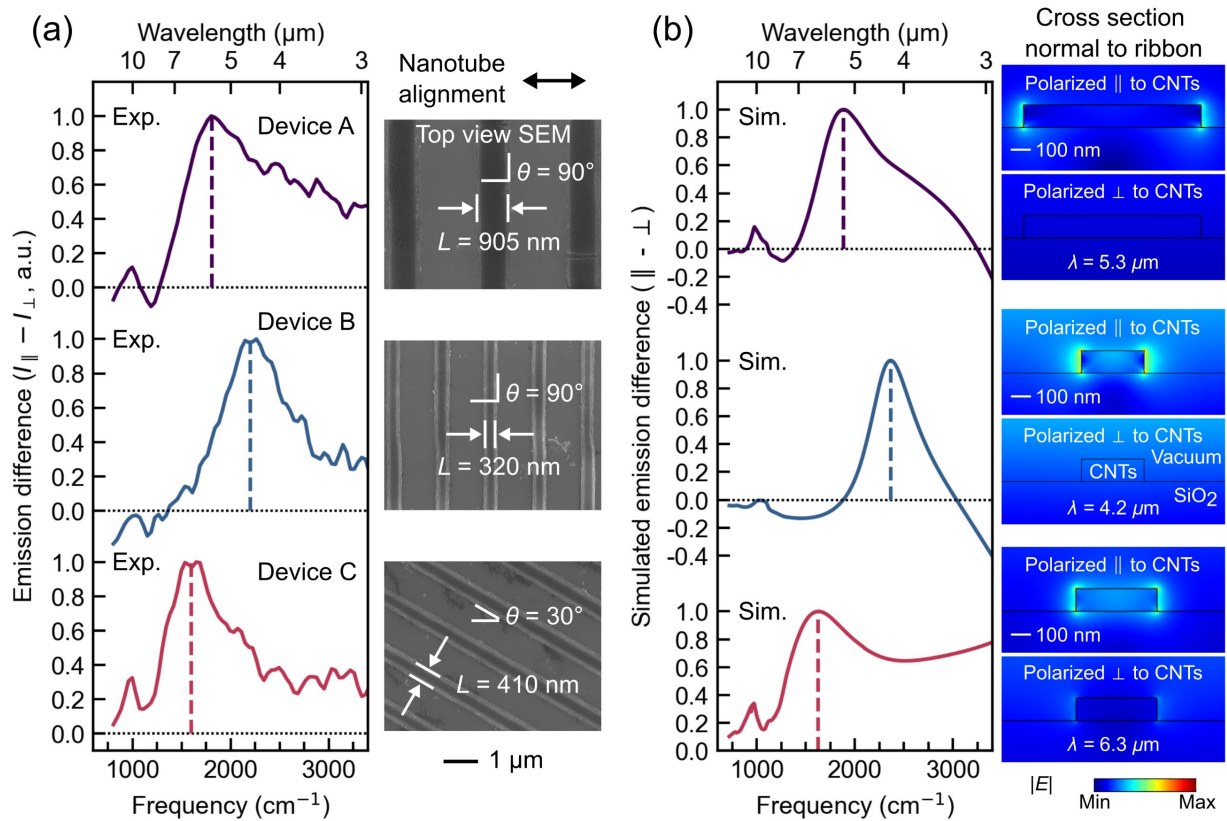

**Figure 2.** Hyperbolic thermal radiators emitting at different peak wavelengths. (a) Polarization difference emission spectra $I_{\parallel}(\omega) - I_{\perp}(\omega)$ of three devices patterned in the same CNT film with varying ribbon geometric parameters. Top: Device A, with relatively wide ribbons ($L = 905$ nm) in a normal-cut direction ($\theta = 90°$), measured at 30 mA. Middle: Device B, with relatively narrow ribbons ($L = 320$ nm) in a normal-cut direction ($\theta = 90°$), at 14 mA (the same data as in Figure 1c). Bottom: Device C, with relatively narrow ribbons ($L = 410$ nm) in an off-axis direction ($\theta = 30°$), at 17.5 mA. Ribbons featuring wider widths and off-normal cut angles show redshifted resonances. (b) Modeled emission spectra corresponding to each device, with ribbon cross-section electric field profiles from finite element optical simulations showing the hyperbolic resonance.

Since the resonance controls the emission spectrum, we expect its shape and peak frequency to be relatively insensitive to the temperature of the resonators. We find experimentally that the peak emission frequency is stable while the ribbon resonators go through large changes in the bias current. Figure 3a shows the emission spectra of Device A at several values of the bias current, showing that the resonant emission frequency is relatively constant as a function of current. The *I-V* curve of Device A (Figure 3b) shows that the current increases rapidly across the range of voltages where we measure the infrared spectrum, corresponding to a significant increase in Joule heating across this range.

The consistent emission peak position and line shape under different bias conditions is remarkable because visible images of the device, when driven with larger currents (Figure 3c), show a large number of apparent hot spots. To investigate the role of hot spots in the resonant infrared emission process, we constructed the apparatus shown in Figure 3d to map the infrared emission intensity across a device by mechanically scanning a single-element detector. Images of the emission from Device A (Figure 3e) indicate that the emission in the mid-wave infrared does originate from hot spots. Additionally, the locations of the infrared hot spots at lower biases are visually correlated with the locations of the visible hot spots at higher biases, and one hot spot does not dominate after the initial 'burn in' of the device (see the Supporting Information).

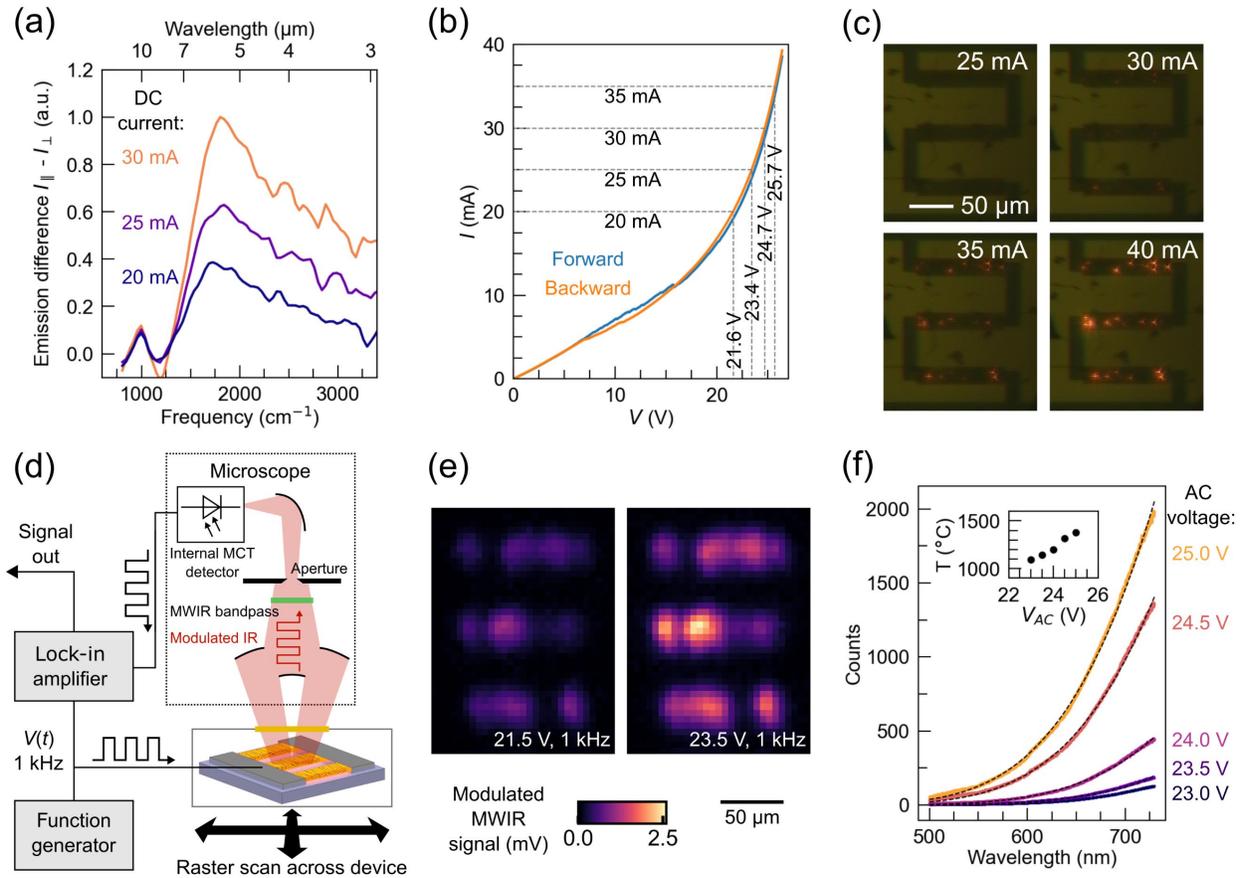

**Figure 3.** Thermal radiation from hot spots. (a) Polarization-difference emission spectra from Device A at several values of the DC current bias. (b) Measurement of the current as a function of bias voltage for Device A with values corresponding to different measurement conditions marked. (c) Optical microscope images of Device A under DC bias at several current levels showing the appearance of visible hot spots and nonuniformity. (d) Schematic of the apparatus used to map the mid-wave infrared emission intensity across the device using a single-element detector in the microscope. The aperture limits the field of view to a small spot. The device bias is modulated with a square wave, and the mid-wave infrared (MWIR) light received by the microscope's single-element detector is measured using a lock-in amplifier. The stage is mechanically scanned to measure the signal at points across the device to form an image. (e) Maps of the MWIR emission intensity from Device A, showing that MWIR emission appears to be correlated with visible hot spots. (f) Visible emission spectra of Device A with fits to Planck's law. Inset: Best-fit values of the temperature for each bias condition.

We estimate the temperature of the hot spots by measuring the spectra of visible emission from the devices. The visible emission spectra of Device A at several square-wave bias voltage levels are shown in Figure 3f. In the visible wavelength range, we expect the CNT metamaterial to have a relatively uniform emissivity spectrum,[36] so that we can estimate a temperature by fitting Planck's law to the visible emission spectrum. The resulting temperatures are shown in the inset of Figure 3f and range roughly from 1090 °C – 1380 °C for biases comparable to those applied during the infrared spectroscopy measurements. Although the precise interpretation of this temperature is difficult due to the nonuniformity of the hot spots, it represents a characteristic temperature for the hottest hot spots that dominate the visible emission and are the largest contributors to the infrared radiation.

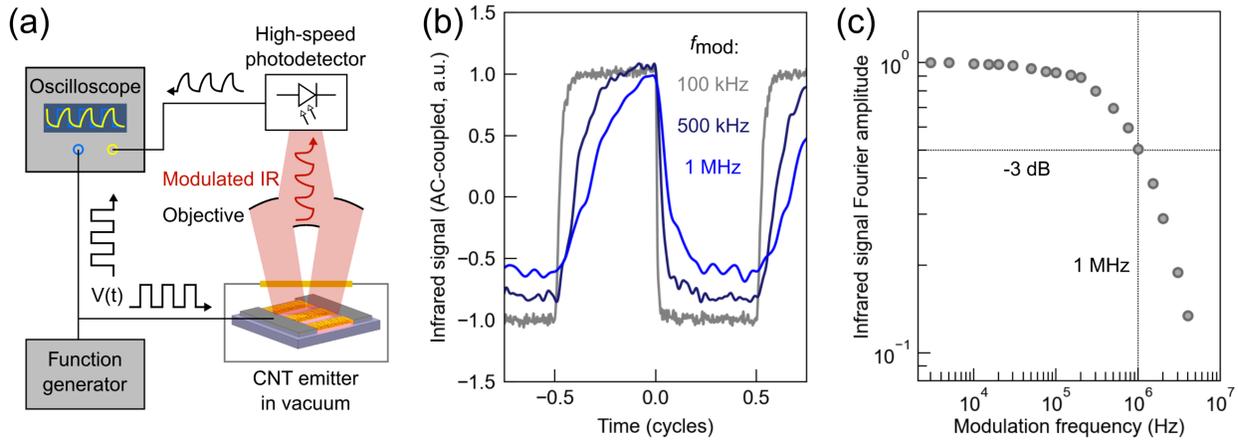

**Figure 4.** Modulation of the thermal radiation up to 1 MHz. (a) Schematic of the modulation measurement setup. An external objective is used to couple light emitted from the device (under vacuum) to a high-speed HgCdTe photodetector. The device is driven using a square wave, and the drive voltage and detector signal are recorded using an oscilloscope. (b) Signal collected from Device A biased using a square wave (3 V – 23 V) at 100 kHz, 500 kHz, and 1 MHz. The signals are digitally low-pass filtered to remove noise. see the Supporting Information for details. (c) Normalized Fourier amplitude of the signal from Device A as a function of the modulation frequency, showing modulation of the thermal radiation at rates up to 1 MHz.

A key feature of the incandescent-resonator device design is that the heated volume is small and in direct contact with the substrate, leading to a short thermal time constant and the potential for modulation much faster than is possible with traditional incandescent emitters. We use a high-speed mid-infrared detector to quantify the response time of our CNT metamaterial emitter (Figure 4a). When operated with a square-wave pulsed drive voltage, the amplitude of the emitted thermal radiation from Device A shows asymmetric rise and fall times on the order of hundreds of nanoseconds (Figure 4b). The maximum modulation rate is ~ 1 MHz measured by the -3 dB frequency of the Fourier component of the infrared emission at the drive voltage (Figure 4c).

This modulation rate corresponds to time constants of the order expected for thermal diffusion in the CNT film. This characteristic timescale[33] is given by $t^2/D$, where $t$ is the CNT film thickness and $D = k/\rho \cdot C_p$ is the thermal diffusivity with thermal conductivity $k$ and volumetric heat capacity $\rho \cdot C_p$. Using the film thickness $t = 115$ nm and a value of $k = 0.085$ W m$^{-1}$ K$^{-1}$ recently reported for a similar CNT film in the perpendicular direction,[58] with a density[59] $\rho = 1.3$ g cm$^{-3}$ and the heat capacity of graphite[60] at 1000 °C, gives a thermal diffusion time constant ~ 400 ns, similar to the experimental result. While these timescales are much longer than for nanoscale metallic films because of the low thermal conductivity of the CNTs in the perpendicular direction, they are orders of magnitude faster than the incandescent sources typically used in commercial infrared sensors. The megahertz modulation bandwidth, together with the ability to batch-fabricate emitters, may enable advances in sensing systems such as high-speed measurements and multi-channel sensing multiplexed by modulation frequency.

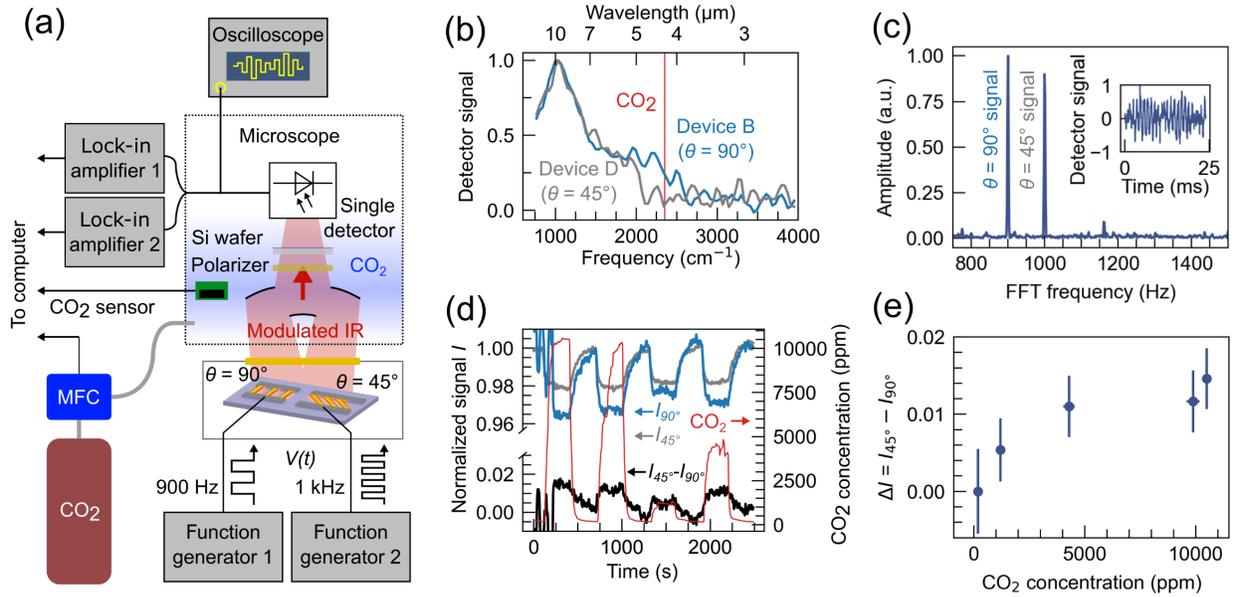

**Figure 5.** Gas sensing with a single detector, using multiple emitters at different emission wavelengths and modulation frequencies. (a) Schematic of the sensing concept. Two devices with different ribbon parameters are simultaneously biased at two different modulation frequencies, and light is collected with one detector. $CO_2$ concentration along the optical path is controlled using a mass flow controller. (b) Measured emission spectra of the two devices at relatively low biases (10 mA for Device B, 8 mA for Device D) for polarization along the CNT axis. Device B has more emission at the $CO_2$ absorption frequency than Device D, where the hyperbolic resonance is redshifted due to the 45° ribbon angle. The spectra are not corrected for the instrument response since the raw detector signal is measured. (c) Detector signal (inset) recorded by the oscilloscope showing a beat pattern, with the Fourier transformed spectrum showing one peak from each device. (d) Amplitudes of the signal from each device (blue and gray) and their difference (black), measured using the lock-in amplifiers and normalized to the levels with no $CO_2$, as the $CO_2$ concentration varies. Device B ($\theta = 90°$) shows larger attenuation with increasing $CO_2$ concentration. (e) Difference between the signals from devices D ($\theta = 45°$) and B ($\theta = 90°$) at different $CO_2$ concentrations based on the data in (d). The difference increases with $CO_2$ concentration. Error bars correspond to the variation in values within the range of times when the $CO_2$ level is considered stable (200-300 s after each flow rate change).

Finally, we demonstrate how two CNT emitters on a single chip, emitting at different wavelengths and with different modulation frequencies, can be used as compact and integrated radiation sources for gas sensing. Our emitter platform is conceptually different from typical nondispersive infrared (NDIR) gas sensors, which use one broadband thermal emitter[3,5] and a pair of detectors with spectral filters tuned to either the gas absorption line or a reference wavelength. Recent work has used spectrally-selective metamaterials to replace the broadband emitter or detector[61–64] but has not incorporated fast modulation or modulation frequency multiplexing. Our emitter concept enables a different system design where multiple spectrally selective emitters, at the gas absorption wavelength and with a detuning, can be received simultaneously by a single broadband detector and distinguished using modulation-frequency multiplexing. While our proof of concept uses a cooled detector, modulation rates in the range of hundreds of Hz can be tailored to maximize the sensitivity of pyroelectric detectors common in NDIR sensors.[65] In general, more emitters on one chip could be operated simultaneously for spectroscopic sensing of broad spectral ranges or detection of multiple gases.

We implement this concept using two CNT emitters on one chip and demonstrate the basic capability to sense $CO_2$ gas. Figure 5a shows a detailed schematic of the measurement. Device B, which has a hyperbolic resonance overlapping the $CO_2$ absorption line[3] at 2349 cm$^{-1}$ (Figure 5b), is biased with a square wave at 900 Hz while an adjacent device (Device D), having the same patterned ribbon width but at an angle $\theta = 45°$ to redshift the hyperbolic resonance away from the $CO_2$ absorption frequency (Figure 5b), is simultaneously biased at 1 kHz for use as a reference. As the $CO_2$ concentration increases, we expect the signal from Device B ($\theta = 90°$) to be attenuated more strongly than that from Device D ($\theta = 45°$). The light emitted from both devices is received by the microscope's internal single-element broadband detector. A portion of the overall signal from the single-element microscope detector with both devices on is shown in the inset of Figure 5c and displays a beat pattern resulting from the two device signals at slightly different frequencies. The Fourier transform of the detector signal shows peaks at the modulation frequencies for each device (Figure 5c).

The amplitudes of these channels (900 Hz sensing and 1 kHz reference), as measured lock-in amplifiers, are shown in Figure 5d as the $CO_2$ concentration inside the microscope is varied using a mass flow controller. Both channels show a decrease in amplitude at higher $CO_2$ concentration (Figure 5d). As expected, Device B ($\theta = 90°$) is attenuated more strongly, and the difference in attenuation between devices B and D increases with $CO_2$ concentration (Figure 5e), indicating the potential to use the pair of devices as a sensor. The key features of the CNT emitter that enable this experiment are the ability to fabricate emitters at different wavelengths side-by-side on the same chip, and to modulate both independently using resistive heating at rates fast enough for frequency multiplexing. To implement a higher-performance sensor it will be necessary to improve the stability and output power of the emitters and to use the full inch-scale CNT films, from which emitter size can be scaled significantly beyond the microscopic footprints of the devices used in this work (330 μm × 180 μm, including contact pads).

In summary, we demonstrate that electrically driven CNT metamaterial ribbons emit spectrally selective thermal radiation at a wavelength determined by the CNT hyperbolic resonance, which can be tuned through the design of the ribbon pattern. The ribbons serve as both efficient resistive heaters and deep-subwavelength hyperbolic resonators, so that the emitters have a compact volume with short thermal time constants that enable emission modulation at rates up to 1 MHz. Multiple emitters with different resonant wavelengths can be fabricated side-by-side on the same chip and operated simultaneously, enabling modulation-frequency-multiplexed gas sensing with multiple wavelength channels. Future work will be needed to understand and mitigate the occurrence of hot spots to improve device uniformity, stability, and output power. This may involve further optimization of the vacuum filtration and CNT film transfer processes. Varying the film preparation parameters will also allow tradeoffs between efficient heating of a thick, thermally insulating CNT film and fast modulation in thin films, as well as tuning of the film's hyperbolic properties through the CNT composition. These design degrees of freedom will complement the tunability of the hyperbolic resonance through the ribbon geometry. We anticipate that the ability to easily integrate emitters tuned to different wavelengths on one chip, with megahertz modulation rates, will enable new system concepts in infrared sensing and communication.

**Author contributions:** J.A.R. and J.F. conceived the experiment. P.H. performed the vacuum filtration process. S.Y. configured the vacuum stage and performed AFM measurements. J.A.R. performed nanofabrication, optical and electrical measurements, analysis, and simulations. All authors wrote and edited the manuscript.

**Acknowledgements:** This work was supported by the Air Force Office of Scientific Research (AFOSR) Multidisciplinary University Research Initiative (MURI) under Award FA9550-16-1-0031 and by the National Science Foundation under Award 2103721. J.A.R. was supported by the Department of Defense through the National Defense Science and Engineering Graduate Fellowship Program. Part of this work was performed at the Stanford Nano Shared Facilities (SNSF) and the Stanford Nanofabrication Facility (SNF), supported by the National Science Foundation under award ECCS-2026822.

**References**

1. Potyrailo, R. A. Multivariable sensors for ubiquitous monitoring of gases in the era of internet of things and industrial internet. *Chem. Rev.* **116**, 11877–11923 (2016).
2. Krier, A. *et al.* Mid-infrared light-emitting diodes. in *Mid-infrared Optoelectronics* (eds. Tournié, E. & Cerutti, L.) 59–90 (Elsevier, 2020). doi:10.1016/B978-0-08-102709-7.00002-4.
3. Popa, D. & Udrea, F. Towards integrated mid-infrared gas sensors. *Sensors* **19**, 2076 (2019).
4. Hildenbrand, J. *et al.* Fast transient temperature operating micromachined emitter for mid-infrared optical gas sensing systems: design, fabrication, characterization and optimization. *Microsyst. Technol.* **16**, 745–754 (2010).

5. Hodgkinson, J. & Tatam, R. P. Optical gas sensing: a review. *Meas. Sci. Technol.* **24**, 012004 (2012).

6. Li, W. & Fan, S. Nanophotonic control of thermal radiation for energy applications [Invited]. *Opt. Express* **26**, 15995–16021 (2018).

7. Baranov, D. G. *et al.* Nanophotonic engineering of far-field thermal emitters. *Nat. Mater.* **18**, 920–930 (2019).

8. Wei, J., Ren, Z. & Lee, C. Metamaterial technologies for miniaturized infrared spectroscopy: Light sources, sensors, filters, detectors, and integration. *J. Appl. Phys.* **128**, 240901 (2020).

9. Puscasu, I. & Schaich, W. L. Narrow-band, tunable infrared emission from arrays of microstrip patches. *Appl. Phys. Lett.* **92**, 233102 (2008).

10. Liu, X. *et al.* Taming the blackbody with infrared metamaterials as selective thermal emitters. *Phys. Rev. Lett.* **107**, 045901 (2011).

11. Liu, J. *et al.* Quasi-coherent thermal emitter based on refractory plasmonic materials. *Opt. Mater. Express* **5**, 2721–2728 (2015).

12. Inoue, T., Zoysa, M. D., Asano, T. & Noda, S. Realization of narrowband thermal emission with optical nanostructures. *Optica* **2**, 27–35 (2015).

13. Shin, S., Elzouka, M., Prasher, R. & Chen, R. Far-field coherent thermal emission from polaritonic resonance in individual anisotropic nanoribbons. *Nat. Commun.* **10**, 1377 (2019).

14. Kudyshev, Z. A., Kildishev, A. V., Shalaev, V. M. & Boltasseva, A. Machine-learning-assisted metasurface design for high-efficiency thermal emitter optimization. *Appl. Phys. Rev.* **7**, 021407 (2020).

15. Suemitsu, M., Asano, T., Inoue, T. & Noda, S. High-efficiency thermophotovoltaic system that employs an emitter based on a silicon rod-type photonic crystal. *ACS Photonics* **7**, 80–87 (2020).

16. He, M. *et al.* Deterministic inverse design of Tamm plasmon thermal emitters with multi-resonant control. *Nat. Mater.* **20**, 1663–1669 (2021).

17. Overvig, A. C., Mann, S. A. & Alù, A. Thermal metasurfaces: complete emission control by combining local and nonlocal light-matter interactions. *Phys. Rev. X* **11**, 021050 (2021).

18. Zhao, B., Song, J.-H., Brongersma, M. & Fan, S. Atomic-scale control of coherent thermal radiation. *ACS Photonics* **8**, 872–878 (2021).

19. Zhou, M. *et al.* Self-focused thermal emission and holography realized by mesoscopic thermal emitters. *ACS Photonics* **8**, 497–504 (2021).

20. Inoue, T., Zoysa, M. D., Asano, T. & Noda, S. Realization of dynamic thermal emission control. *Nat. Mater.* **13**, 928–931 (2014).

21. Brar, V. W. *et al.* Electronic modulation of infrared radiation in graphene plasmonic resonators. *Nat. Commun.* **6**, 7032 (2015).

22. Xiao, Y., Charipar, N. A., Salman, J., Piqué, A. & Kats, M. A. Nanosecond mid-infrared pulse generation via modulated thermal emissivity. *Light Sci. Appl.* **8**, 51 (2019).

23. Song, B. *et al.* Giant gate-tunability of complex refractive index in semiconducting carbon nanotubes. *ACS Photonics* **7**, 2896–2905 (2020).

24. Zou, Y. *et al.* Non-Planckian infrared emission from GaAs devices with electrons and lattice out-of-thermal-equilibrium. *Opt. Express* **29**, 1244–1250 (2021).

25. Pyatkov, F. *et al.* Cavity-enhanced light emission from electrically driven carbon nanotubes. *Nat. Photonics* **10**, 420–427 (2016).

26. Miyoshi, Y. *et al.* High-speed and on-chip graphene blackbody emitters for optical communications by remote heat transfer. *Nat. Commun.* **9**, 1279 (2018).

27. Shi, C., Mahlmeister, N. H., Luxmoore, I. J. & Nash, G. R. Metamaterial-based graphene thermal emitter. *Nano Res.* **11**, 3567–3573 (2018).

28. Kim, Y. D. *et al.* Ultrafast graphene light emitters. *Nano Lett.* **18**, 934–940 (2018).

29. Brouillet, J., Papadakis, G. T. & Atwater, and H. A. Experimental demonstration of tunable graphene-polaritonic hyperbolic metamaterial. *Opt. Express* **27**, 30225–30232 (2019).

30. Kocer, H. *et al.* Exceptional adaptable MWIR thermal emission for ordinary objects covered with thin $VO_2$ film. *J. Quant. Spectrosc. Radiat. Transfer* **262**, 107500 (2021).

31. Mori, T., Yamauchi, Y., Honda, S. & Maki, H. An electrically driven, ultrahigh-speed, on-chip light emitter based on carbon nanotubes. *Nano Lett.* **14**, 3277–3283 (2014).

32. Sakat, E. *et al.* Enhancing thermal radiation with nanoantennas to create infrared sources with high modulation rates. *Optica* **5**, 175–179 (2018).

33. Wojszvzyk, L. *et al.* An incandescent metasurface for quasimonochromatic polarized mid-wave infrared emission modulated beyond 10 MHz. *Nat Commun* **12**, 1492 (2021).

34. Fujiwara, M., Tsuya, D. & Maki, H. Electrically driven, narrow-linewidth blackbody emission from carbon nanotube microcavity devices. *Appl. Phys. Lett.* **103**, 143122 (2013).

35. Roberts, J. A. *et al.* Tunable hyperbolic metamaterials based on self-assembled carbon nanotubes. *Nano Lett.* **19**, 3131–3137 (2019).

36. Schöche, S. *et al.* Mid-IR and UV-Vis-NIR Mueller matrix ellipsometry characterization of tunable hyperbolic metamaterials based on self-assembled carbon nanotubes. *J. Vac. Sci. Technol. B* **38**, 014015 (2020).

37. He, X. *et al.* Wafer-scale monodomain films of spontaneously aligned single-walled carbon nanotubes. *Nat. Nanotechnol.* **11**, 633–638 (2016).

38. Gao, W. *et al.* Macroscopically aligned carbon nanotubes for flexible and high-temperature electronics, optoelectronics, and thermoelectrics. *J. Phys. D: Appl. Phys.* **53**, 063001 (2019).

39. Tulevski, G. S. & Falk, A. L. Emergent properties of macroscale assemblies of carbon nanotubes. *Adv. Funct. Mater.* **30**, 1909448 (2020).

40. Chiu, K.-C. *et al.* Strong and broadly tunable plasmon resonances in thick films of aligned carbon nanotubes. *Nano Lett.* **17**, 5641–5645 (2017).

41. Ho, P.-H. *et al.* Intrinsically ultrastrong plasmon–exciton interactions in crystallized films of carbon nanotubes. *Proc. Natl. Acad. Sci. U.S.A.* **115**, 12662–12667 (2018).

42. Gao, W., Doiron, C. F., Li, X., Kono, J. & Naik, G. V. Macroscopically aligned carbon nanotubes as a refractory platform for hyperbolic thermal emitters. *ACS Photonics* **6**, 1602–1609 (2019).

43. Roberts, J. A. *et al.* Multiple tunable hyperbolic resonances in broadband infrared carbon-nanotube metamaterials. *Phys. Rev. Applied* **14**, 044006 (2020).

44. Adhikari, C. M. & Bondarev, I. V. Controlled exciton–plasmon coupling in a mixture of ultrathin periodically aligned single-wall carbon nanotube arrays. *J. Appl. Phys.* **129**, 015301 (2021).

45. Bondarev, I. V. & Adhikari, C. M. Collective excitations and optical response of ultrathin carbon-nanotube films. *Phys. Rev. Applied* **15**, 034001 (2021).

46. Sveningsson, M., Jönsson, M., Nerushev, O. A., Rohmund, F. & Campbell, E. E. B. Blackbody radiation from resistively heated multiwalled carbon nanotubes during field emission. *Appl. Phys. Lett.* **81**, 1095–1097 (2002).

47. Li, P. *et al.* Polarized incandescent light emission from carbon nanotubes. *Appl. Phys. Lett.* **82**, 1763–1765 (2003).

48. Mann, D. *et al.* Electrically driven thermal light emission from individual single-walled carbon nanotubes. *Nat. Nanotechnol.* **2**, 33–38 (2007).

49. Liu, P. *et al.* Fast high-temperature response of carbon nanotube film and its application as an incandescent display. *Adv. Mater.* **21**, 3563–3566 (2009).

50. Liu, Z., Bushmaker, A., Aykol, M. & Cronin, S. B. Thermal emission spectra from individual suspended carbon nanotubes. *ACS Nano* **5**, 4634–4640 (2011).

51. Bao, W. *et al.* Flexible, high temperature, planar lighting with large scale printable nanocarbon paper. *Adv. Mater.* **28**, 4684–4691 (2016).

52. Desai, P. D., Chu, T. K., James, H. M. & Ho, C. Y. Electrical resistivity of selected elements. *J. Phys. Chem. Ref. Data* **13**, 1069–1096 (1984).

53. Griffiths, P. R. & De Haseth, J. A. *Fourier Transform Infrared Spectrometry*. (John Wiley & Sons, Inc., 2007).

54. Xiao, Y. *et al.* Measuring thermal emission near room temperature using Fourier-transform infrared spectroscopy. *Phys. Rev. Applied* **11**, 014026 (2019).

55. Xiao, Y. *et al.* Precision measurements of temperature-dependent and nonequilibrium thermal emitters. *Laser Photonics Rev.* **14**, 1900443 (2020).

56. Falk, A. L. *et al.* Coherent plasmon and phonon-plasmon resonances in carbon nanotubes. *Phys. Rev. Lett.* **118**, 257401 (2017).

57. Greffet, J.-J., Bouchon, P., Brucoli, G. & Marquier, F. Light emission by nonequilibrium bodies: Local Kirchhoff law. *Phys. Rev. X* **8**, 021008 (2018).

58. Yamaguchi, S. *et al.* One-directional thermal transport in densely aligned single-wall carbon nanotube films. *Appl. Phys. Lett.* **115**, 223104 (2019).

59. Behabtu, N. *et al.* Strong, light, multifunctional fibers of carbon nanotubes with ultrahigh conductivity. *Science* **339**, 182–186 (2013).

60. Hone, J. *et al.* Thermal properties of carbon nanotubes and nanotube-based materials. *Appl. Phys. A* **74**, 339–343 (2002).

61. Lochbaum, A. *et al.* Compact mid-infrared gas sensing enabled by an all-metamaterial design. *Nano Lett.* **20**, 4169–4176 (2020).

62. Livingood, A. *et al.* Filterless nondispersive infrared sensing using narrowband infrared emitting metamaterials. *ACS Photonics* **8**, 472–480 (2021).

63. Pusch, A. *et al.* A highly efficient CMOS nanoplasmonic crystal enhanced slow-wave thermal emitter improves infrared gas-sensing devices. *Sci. Rep.* **5**, 17451 (2015).

64. Hasan, D. & Lee, C. Hybrid metamaterial absorber platform for sensing of $CO_2$ gas at mid-IR. *Adv. Sci.* **5**, 1700581 (2018).

65. Rogalski, A. *Infrared Detectors*. (CRC Press, 2000).